\documentclass[preprint,aps,prl,superscriptaddress,floatfix,longbibliography]{revtex4-2}
\usepackage{amsmath,amssymb}
\usepackage{mathtools}
\usepackage{graphicx}
\usepackage[colorlinks=true,linkcolor=blue,citecolor=blue,urlcolor=blue]{hyperref}
\usepackage[mathlines]{lineno}
\newcommand{\eqdef}{\coloneqq}
\graphicspath{{figures/}{../figures/}}

\begin{document}
%\nocite{apsrev4-2Control}
%\linenumbers

\title{Linear Response Predicts Cusp-Pair Births in Networks with a Localized Cubic}
\author{Kanchan Sarkar}
\affiliation{Institut f\"ur Theoretische Chemie, Universit\"at Ulm, 89081 Ulm, Germany}
\date{\today}

\begin{abstract}
Linear response is cheap to measure; the bistability boundaries it organizes are not. For a passive network with one localized cubic, the driving-point receptance $G$ fixes the period-one cusp set at fundamental-harmonic order: cusps lie on a fixed phase contour of $G$, a tangency of that contour under parameter variation creates a pair, and its curvature separates a gap opening from an isolated loop. For a two-mode absorber the linear prediction locates a benchmark birth coupling to $0.3\%$, and to $0.03\%$ once a third-harmonic correction of scale $|G(3\Omega)/G(\Omega)|$ is included.
\end{abstract}

\maketitle

Linear frequency response is often available from a small-signal sweep or a
linear model. Mapping the forced response of a nonlinear network usually
requires sweeps in both drive and frequency. We ask how much of the period-one fold set is already fixed by the linear response. That set comprises the turning points of the periodic response and the cusps at which those turning points reconnect.

\begin{figure*}[!tbp]
  \centering
  \includegraphics[width=\textwidth]{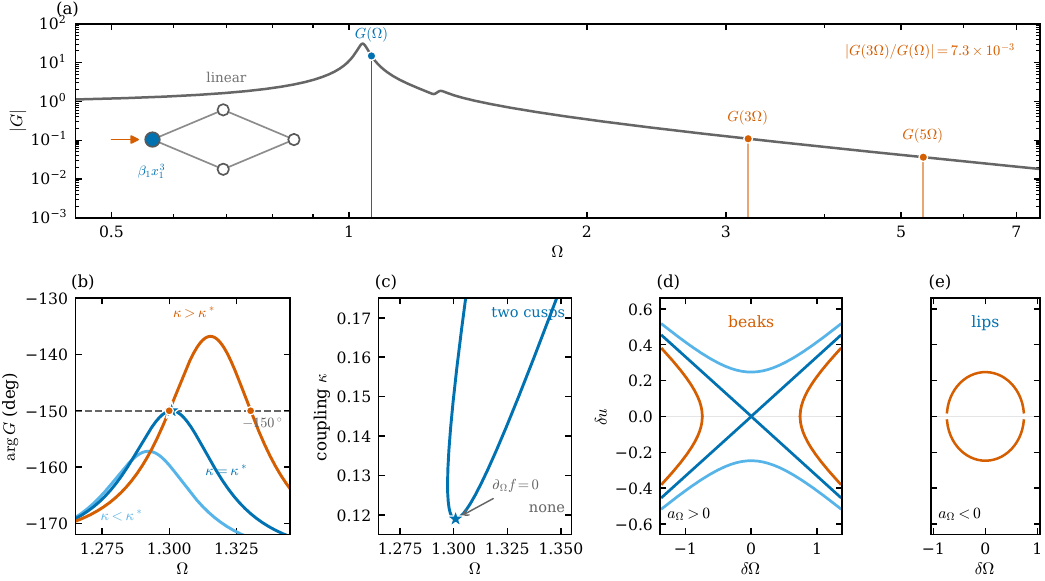}
  \caption{The construction. (a)~A network with one nonlinear coordinate
    reduces to the driving-point receptance $G$ at that coordinate, sampled at
    the harmonic frequencies. In the two-mode absorber,
    $|G(3\Omega)/G(\Omega)|$ is $7.3\times10^{-3}$ at the benchmark cusp
    ($\kappa=0.10$) and $4.2\times10^{-2}$ at the benchmark birth shown in
    (b)--(e) ($\kappa^\ast=0.1190$), i.e.\ $\eta_3=1.4\times10^{-3}$ and
    $8.2\times10^{-3}$; the birth lies between the antiresonance at $1.2967$
    and the modal frequency $1.3060$.
    (b)~Below the birth coupling, $\arg G$ does not reach $-150^\circ$; at
    $\kappa^\ast$ it is tangent to that level; above it, the two crossings are
    the two cusps. (c)~In the $(\Omega,\kappa)$ plane, the same phase level is
    the cusp locus. Its fold under projection onto $\kappa$ satisfies
    $\partial_\Omega f=0$. (d),(e)~Local fold sets from
    $9\delta u^2-a_\Omega\delta\Omega^2=b_\kappa\delta\kappa$, at three values
    of $b_\kappa\delta\kappa$ coloured as in (b): for $a_\Omega>0$ the branches
    reconnect, leaving a gap; for $a_\Omega<0$ they form an isolated oval.}
  \label{fig:construction}
\end{figure*}

For a passive network whose cubic acts on a single coordinate, and once the cubic form and its sign are known, the small-signal driving-point receptance $G$ at that coordinate answers this at
fundamental-harmonic order. Cusps lie on a fixed phase
contour of $G$. Where the contour folds under projection onto a varied network
parameter, a cusp pair is created. The curvature at the fold decides whether
the fold set opens a gap or closes into an isolated loop
(Fig.~\ref{fig:construction}). The prediction requires no nonlinear response surface, and the same conditions can be solved backward to place a birth at a prescribed frequency.

The same receptance also sets the scale of the leading third-harmonic
correction. At a cusp the cubic magnitude cancels from that correction relative
to the fundamental balance, leaving a scale fixed by a ratio of linear response
at two frequencies, $|G(3\Omega)/G(\Omega)|$. In the contour tests below the located cusps reach an anharmonicity of $0.79$ while $|G(3\Omega)/G(\Omega)|$ stays below $0.08$, so the construction remains accurate where $\varepsilon$ is no longer small. Branch stability remains a separate
Floquet question.

The Duffing oscillator supplies the familiar special case. At
fundamental-harmonic order, bistability begins when the detuning exceeds
$\sqrt3$ damping half-widths~\cite{HolmesRand1976,LifshitzCross2008}, a
threshold familiar in nanomechanics~\cite{AldridgeCleland2005,AlmogEtAl2007}
and Kerr resonators~\cite{RobertsClerk2020,TritschlerEtAl2025}.
Describing-function theory gives that condition a geometric form: sinusoidal
balance for a cubic around a linear plant predicts a jump when the linear
frequency-response locus enters a region fixed by the nonlinearity. The cubic
construction dates to the 1950s~\cite{WestDouceLivesley1956}, and later work
expressed jump conditions through complex-plane loci and graphical criteria for
higher-order nonlinear feedback systems~%
\cite{Hatanaka1963,FukumaMatsubara1966,MurtyKrishnaDeekshatulu1968,FukumaMatsubara1978PartI,HiraiSawai1978}.
The fixed boundary is classical, and for an arbitrary linear plant, not only a
single Lorentzian: sinusoidal balance for a cubic gives the discriminant
$\rho^2=3\sigma^2$, so Eq.~\eqref{eq:phasecontour} restates the jump-resonance
criterion of that literature in phase form. What is new is what that boundary
does when a network parameter is varied. Read as a level set in
$(\Omega,\kappa)$, its tangency locates pair creation and its curvature
classifies the reconnection.

Closely related two-mode linear--Duffing geometries have been studied
directly, with detached response curves appearing and disappearing as system
parameters
change~\cite{GattiBrennanKovacic2010,GattiKovacicBrennan2010,GattiBrennan2011,Gatti2016},
one of which has been built and measured~\cite{GattiBrennan2017}. There the
cubic acts on a relative coordinate and the drive on the linear mass: a
noncollocated member of the class treated here, whose nonlinear response those
studies construct explicitly. The condition used here is imposed on the linear
response alone and holds for any passive network with a localized cubic.
Bifurcation tracking in harmonic balance reaches the same folds and cusps by
continuing the nonlinear response surface~\cite{DetrouxEtAl2015}; the input
here is one small-signal sweep.

Island-like jump responses were identified within describing-function theory,
and modern variants design multi-jump and multiple-hysteresis frequency
responses~%
\cite{GelbVanderVelde1968,Atherton1975,KoenigsbergDunn1975,BuscarinoFamosoFortunaFrasca2020,BucoloBuscarinoFortunaFrasca2020,BuscarinoFamosoFortuna2025,MarinoSodja2026};
singularity and reduced-order methods locate and design detached responses from
nonlinear frequency-response equations, reduced dynamics, or bifurcation
tracking~%
\cite{Cirillo2017,HabibCirilloKerschen2018,Ponsioen2019,MelotDenimalRenson2024,DekemeleHabib2025}.
A complementary line classifies the stationary-state topology of
driven-dissipative nonlinear systems directly~\cite{VillaEtAl2025}.
Closest to the present construction, Kyzio{\l} and Okni{\'n}ski derive an
effective nonlinear equation for coupled driven oscillators and classify the
metamorphoses of its amplitude profiles through the singular points of the
bifurcation
set~\cite{KyziolOkninski2011,KyziolOkninski2017,KyziolOkninski2020}. That route builds the nonlinear response first and then examines its singularities, whereas the organizer here is the small-signal receptance of an otherwise arbitrary linear environment.

\textit{Reduction to the local receptance.}---Consider an
$N$-degree-of-freedom linear network with real symmetric mass, damping, and
stiffness matrices $M$, $C$, and $K$, with $C\succeq0$, so that
$Z(\Omega)=K-\Omega^2M+i\Omega C$. Let the coordinate selected by $e_1$ carry
both the drive and a cubic nonlinearity $\beta_1x_1^3$. Expanding the periodic
steady state as $x(t)=\sum_nX^{(n)}e^{in\Omega t}$, balance at harmonic $n$
reads
\begin{equation}
Z(n\Omega)\,X^{(n)} + \beta_1 (x_1^3)^{(n)} e_1
   = \tfrac{F}{2}\bigl(\delta_{n,1} + \delta_{n,-1}\bigr) e_1.
\label{eq:combbalance}
\end{equation}
Because both the forcing and the nonlinear term act along $e_1$, solving the
linear network and projecting back onto $e_1$ gives a scalar balance at each
harmonic,
\begin{equation}
\begin{aligned}
X_1^{(n)} &= G(n\Omega)\Bigl[\tfrac{F}{2}
   \bigl(\delta_{n,1}+\delta_{n,-1}\bigr) - \beta_1 (x_1^3)^{(n)}\Bigr],\\
G &\eqdef e_1^{T} Z^{-1} e_1.
\end{aligned}
\label{eq:combclosure}
\end{equation}
No harmonic has been truncated in Eq.~\eqref{eq:combclosure}. The entire linear
network enters the scalar balance through the driving-point receptance sampled
at $n\Omega$. This condensation is standard for localized
nonlinearities~\cite{RenBeards1994}; in circuit QED, the analogous object is the
impedance presented by a linear environment to a Josephson element~%
\cite{NiggEtAl2012}. For fixed $F$ and $\beta_1$, two networks with the same
sampled receptance comb therefore obey the same periodic-balance equation at
the nonlinear coordinate. Extensions to a noncollocated drive and more general
linear damping, together with the viscous Floquet closure, are given in End
Matter.

\textit{Cusps are a phase contour.}---Equation~\eqref{eq:combclosure} is
exact; the fundamental-harmonic truncation enters here and nowhere else. Keeping
only $n=1$, define the physical harmonic amplitudes
$A_n\eqdef2X_1^{(n)}$ by
$x_1=\operatorname{Re}\sum_{n>0}A_ne^{in\Omega t}$, and write
$1/G=\rho+i\sigma$ and $y=|A_1|^2$. The amplitude equation is
$F^2=y[(\rho+\tfrac34\beta_1y)^2+\sigma^2]$. At fixed $\Omega$, folds are the
turning points of this drive--amplitude relation, so
$\partial_yF^2=0$. A cusp occurs when the two fold roots merge. Setting the
quadratic discriminant to zero gives $\rho^2=3\sigma^2$, while a positive
physical amplitude requires $\beta_1\rho<0$. Together these conditions fix the
cusp phase:
\begin{equation}
\arg G(\Omega_c) =
\begin{cases}
-150^\circ, & \beta_1 > 0 \quad(\text{hardening}),\\[2pt]
-30^\circ,  & \beta_1 < 0 \quad(\text{softening}),
\end{cases}
\label{eq:phasecontour}
\end{equation}
for physical cusps with $\Omega_c>0$. The conjugate levels $+150^\circ$ and
$+30^\circ$ are inaccessible in a passive collocated network: with
$C\succeq0$, $\operatorname{Im}G\le0$ for $\Omega>0$, so
$\arg G\in[-180^\circ,0^\circ]$ (End Matter). Both cusps of a born pair lie on
the same physical contour within the fundamental-harmonic reduction
(Fig.~\ref{fig:construction}b).

\begin{figure*}[!tbp]
  \centering
  \includegraphics[width=\textwidth]{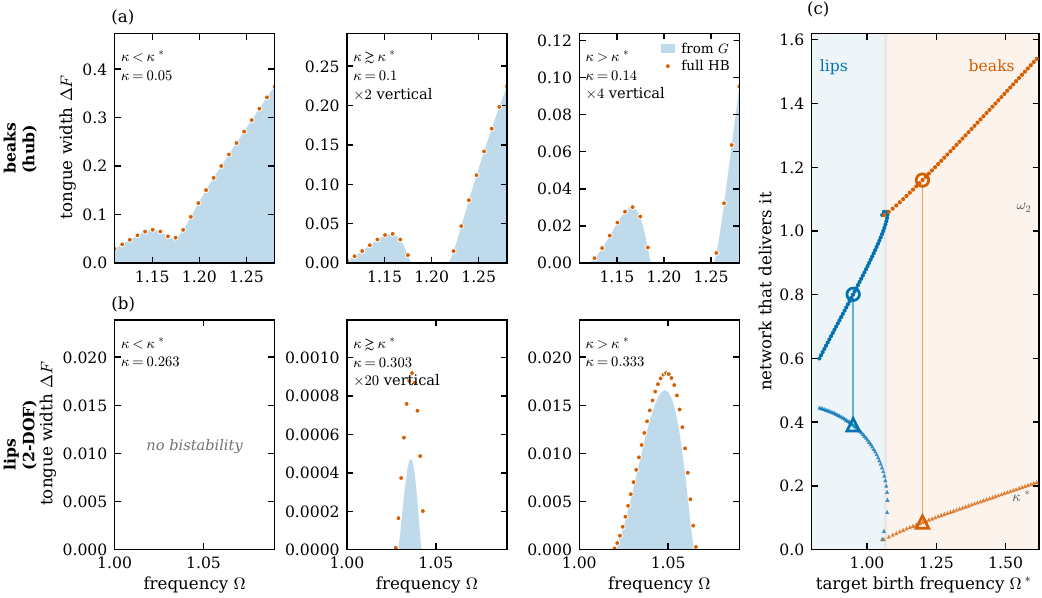}
  \caption{Linear response as a design variable. (a),(b)~Bistable tongue width
    $\Delta F$ versus drive frequency below, near, and above a birth. Shading
    is the fundamental-harmonic result from $G$ and $\beta_1$; points are from
    multi-harmonic balance of the original equations, at five harmonics for the
    hub and seven for the absorber. Vertical scales are per panel, tagged
    where magnified against the widest panel in the row. (a)~A three-resonator hub
    ($\omega=1,1.15,1.30$) develops a gap bounded by the born cusp pair, a
    \emph{beaks}-type birth. (b)~With the absorber's auxiliary mode below the
    drive ($\omega_2=0.95$), an isolated bistable lens appears, a
    \emph{lips}-type birth; at $\kappa=0.303$ its cusps have
    $\varepsilon=0.29$ and $0.42$, and the resolved lens is wider than the
    fundamental-harmonic result while its topology is unchanged. (c)~Inverse
    design from \eqref{eq:birth}, sampled at $99$ births. For each target
    $\Omega^\ast$, circles and triangles give the resulting $\omega_2$ and
    $\kappa^\ast$; shading gives the type from \eqref{eq:aomega}; open symbols
    mark the two designs carried into harmonic balance. Both types are
    available for $\Omega^\ast\in[1.061,1.074]$, where the birth locus folds
    and $a_\Omega$ changes sign.}
  \label{fig:design}
\end{figure*}

We use two systems for numerical tests. The first is a two-degree-of-freedom
Duffing absorber with the cubic on the driven coordinate,
\begin{align}
\ddot{x}_1 + 2\gamma_1\dot{x}_1 + \omega_1^2 x_1
  + \kappa(x_1 - x_2) + \beta_1 x_1^3 &= F\cos\Omega t, \nonumber\\
\ddot{x}_2 + 2\gamma_2\dot{x}_2 + \omega_2^2 x_2
  + \kappa(x_2 - x_1) &= 0,
\end{align}
with $(\omega_1,\omega_2) = (1,\,1.25)$,
$(\gamma_1,\gamma_2) = (0.015,\,0.02)$ and $\beta_1 = 0.147$ unless stated
otherwise; each $\gamma_j$ is a damping rate, a half-width, not a
dimensionless ratio. The second is the
three-resonator hub of Fig.~\ref{fig:design}a. The conditions derived below
depend only on $G$.

Two dimensionless quantities recur in the tests. The anharmonicity
$\varepsilon\eqdef|\beta_1|\,|A_1|^2/\omega_1^2$ measures the cubic stiffness
shift relative to $\omega_1^2$ and provides the small parameter for a generic
weak-anharmonicity expansion away from a cusp. At a fundamental cusp, the
double-root condition gives
$\varepsilon=\tfrac89|\rho|/\omega_1^2$. The leading third-harmonic feedback
has a different scale. At the cusp the double root pins $\tfrac34\beta_1|A_1|^2=-\tfrac23\rho$, so the first correction
relative to the cubic balance is
\begin{equation}
\eta_3 = \frac{\sqrt3}{9}\left|\frac{G(3\Omega)}{G(\Omega)}\right|,
\label{eq:eta3}
\end{equation}
which depends only on the linear receptance, the same small-signal data that
locate the cusp~\cite{SM}. It need not track $\varepsilon$; the stress test below sweeps $\eta_3$ twentyfold at fixed $\varepsilon$.

The rescaling $\tilde x=\sqrt{|\beta_1|}\,x$, $\tilde F=\sqrt{|\beta_1|}\,F$
removes the cubic magnitude from the full equations, which is why it is absent
from \eqref{eq:eta3}: changing $|\beta_1|$ rescales amplitudes and leaves every
frequency coordinate, cusps included, where it was.

\textit{Testing the phase contour.}---We compare \eqref{eq:phasecontour} with cusps
located independently using multi-harmonic balance of the original equations.
Their frequencies come from the $(\Delta F)^{2/3}$ tongue-width intercept,
which neither solves at the degenerate tip nor uses the phase condition~%
\cite{SM}. For $28$ cusps of the two-mode absorber, spanning
$\Omega=1.30$--$1.43$ as the coupling changes by a factor of $2.5$, the phase
of $G$ at the located cusps stays within $0.44^\circ$ of $-150^\circ$. Over
that band $\arg G$ itself sweeps $27^\circ$ to $92^\circ$, depending on
coupling, so the residual is under $2\%$ of the range traversed. Across those
cusps $\varepsilon$ runs from $0.27$ to $0.79$; four further cusps reaching
$\varepsilon=1.18$ are predicted but not resolved by the estimator and are
excluded~\cite{SM}. Evaluated
separately at each cusp, the first comb correction has the same sign as the
observed frequency residual at all $28$ cusps and lowers the median discrepancy
from $2.0\times10^{-4}$ to $1.0\times10^{-4}$. The softening branch gives a
separate check: with $\beta_1=-0.147$, eight cusps lie within $0.047^\circ$
of $-30^\circ$, and their smaller residual is comparable with the estimator
bias. The full contour test and error budget are given in the Supplement~%
\cite{SM}.

\textit{Cusp-pair births.}---In the plane of drive frequency and coupling,
Eq.~\eqref{eq:phasecontour} is a level set. A pair is born when this level set turns back in $\Omega$ at fixed
coupling, equivalently when it folds under projection onto the coupling
(Fig.~\ref{fig:construction}c):
\begin{equation}
f = 0, \qquad \partial_\Omega f = 0,
\qquad f \eqdef \rho^2 - 3\sigma^2.
\label{eq:birth}
\end{equation}
The birth is generic when $\partial_\kappa f\neq0$ and
$\partial_\Omega^2f\neq0$. Both conditions involve only $G$. Additional
resonances visible at the nonlinear coordinate can generate additional folds of
the contour, and hence additional births; the companion paper resolves this
mechanism pole by pole~\cite{Sarkar2026Long}.

\textit{Local birth type.}---The scalar $f$ also determines the
local geometry. Define the curvature coefficient at a birth by
\begin{equation}
a_\Omega \eqdef \tfrac12\,\partial_\Omega^2 f.
\label{eq:aomega}
\end{equation}
The classifier uses the full receptance and needs no pole approximation. To
see its role, let $u\eqdef\rho+\tfrac34\beta_1|A_1|^2$. Completing the square in the fold
equation gives
$9(u-\rho/3)^2=f$, so $f>0$ gives two fold roots in $u$ and $f<0$ gives
none. Near a birth, set $\delta u\eqdef u-\rho/3$,
$\delta\Omega=\Omega-\Omega^\ast$, and
$\delta\kappa=\kappa-\kappa^\ast$. Since $f=\partial_\Omega f=0$ at the
birth,
$f=a_\Omega\delta\Omega^2+b_\kappa\delta\kappa+\cdots$, where
$b_\kappa\eqdef(\partial_\kappa f)_\ast$. The local fold set is therefore
\[
9\delta u^2-a_\Omega\delta\Omega^2=b_\kappa\delta\kappa,
\]
to leading order. The sign of $b_\kappa$ selects the side of the threshold on which the folds
exist; the sign of $a_\Omega$ determines how they meet. The resulting conics
match the standard plane-to-plane beaks/lips geometries~%
\cite{Arnold1981,Saji2010,Kuznetsov2004}: $a_\Omega>0$ gives a
\emph{beaks}-type gap-opening reconnection, whereas $a_\Omega<0$ gives a
\emph{lips}-type isolated-loop birth.

\textit{Beyond the fundamental.}---A cubic response dominated by the
fundamental also generates odd harmonics, led by $3\Omega$. Higher-order
sinusoidal-input describing functions describe such responses
generally~\cite{NuijBosgraSteinbuch2006}, and higher-harmonic feedback can be
structurally decisive for other singular objects~\cite{Sarkar2026PartI}. Here
the cusp double-root condition gives a particularly simple first feedback
scale. To
first order in the generated third harmonic, the $n=3$ balance in
\eqref{eq:combclosure} gives $A_3=-\tfrac14\beta_1G_3A_1^3$, where
$G_3\eqdef G(3\Omega)$. Feeding this term back into the fundamental balance and
retaining the first such contribution gives, with $G_1\eqdef G(\Omega)$,
\begin{equation}
F = A_1\Bigl[\frac{1}{G_1} + \tfrac34\beta_1|A_1|^2
    - \tfrac{3}{16}\beta_1^2\,G_3\,|A_1|^4\Bigr].
\label{eq:quintic}
\end{equation}
At a fundamental cusp the amplitude rescales with the cubic coefficient, so the
relative size of this feedback reduces to Eq.~\eqref{eq:eta3}; the cancellation
is derived in the Supplement. At the benchmark cusp the corrected condition
shifts $\Omega_c$ by $-2.4\times10^{-5}$, below the resolution of the
fold-tracking estimator.

\textit{Separating comb gain from anharmonicity.}---A network with a resonance
deliberately placed near $3\Omega$ separates the two parameters. Three modes, the first two fixing a cusp at $\varepsilon=1.03$ and
the third tuned through $3\Omega_c$, sweep $\eta_3$ from $0.005$ to $0.103$
while $\varepsilon$ moves by less than $0.1\%$. Against converged
multi-harmonic balance, the fundamental condition holds to a median
$7\times10^{-4}$ in $\Omega$ over the eight weakest-comb members. At the two
largest gains, $\eta_3=0.085$ and $0.103$, it misplaces the cusp by
$8\times10^{-3}$ and $4.5\times10^{-3}$, at the first of these predicting a
bistable band in which the multi-harmonic solution has none;
Eq.~\eqref{eq:quintic} pulls both back to about $10^{-3}$~\cite{SM}. Within this family the comb gain sets the scale on which the construction degrades. It does not order the individual residuals: $G(3\Omega)$ enters the corrected fold condition through its real and imaginary parts separately, whereas $\eta_3$ is a modulus, so the signed shift is not fixed by $\eta_3$ alone~\cite{SM}.

The type survives that correction. Carrying \eqref{eq:quintic} through the fold
condition gives $\delta a_\Omega$ at the same order, and the classification is
unchanged while $|\delta a_\Omega|<|a_\Omega|$. Away from the type boundary in
Fig.~\ref{fig:design}c, $|\delta a_\Omega|/|a_\Omega|$ runs from $0.7\%$ to
$2.8\%$ and tracks the comb ratio, growing as $3\Omega$ nears a resonance or
$a_\Omega$ nears zero. The window about $a_\Omega=0$ in which the correction
could overturn the classification is narrow, estimated at
$\sim\!2\times10^{-6}$ in $\kappa$ for the two-mode family (End Matter).

\textit{Validation.}---We locate the birth of the two-mode absorber without using either cusp condition. In the harmonic-balance solution we
bisect on the coupling at which a gap first opens in the bistable band: a
topological verdict, using neither \eqref{eq:phasecontour} nor the $3/2$
tongue-width law. Two independently configured runs of that test bracket
$\kappa^\ast$ in $[0.1193188,0.1193203]$ (End Matter). The verdict is
unchanged from three harmonics to nine, so the bracket is not set by the
truncation. Solving \eqref{eq:birth} on the linear receptance gives
$(\Omega^\ast,\kappa^\ast)=(1.30088,0.11896)$, which misses the bracket
midpoint by $0.301\%$ at a birth where $\varepsilon=0.46$; the first
third-harmonic correction moves it to $(1.30105,0.11929)$, which misses by
$0.026(1)\%$ (Supplement).

Multi-harmonic balance also reproduces both local reorganizations in
Fig.~\ref{fig:design}a,b: in the hub, a connected bistable band develops a gap
at the predicted cusp pair; in the absorber, an isolated lens appears above the
birth and none below, its resolved edges agreeing with the predicted lens to
within the scan resolution.
The located tips again come from $(\Delta F)^{2/3}$ fits with the generic
$3/2$ exponent and no fitted prefactor. Direct monodromy integration confirms Floquet-stable coexisting outer branches
in both examples (End Matter).

\textit{Design corollary.}---Given the small-signal receptance $G$, choose the
physical phase branch in \eqref{eq:phasecontour} and solve \eqref{eq:birth} as
the network parameter varies. Retain solutions that satisfy the physical sign
and genericity conditions; \eqref{eq:aomega} gives the local type. Conversely,
\eqref{eq:birth} solves for network parameters that place a birth at a
prescribed frequency. In the two-mode family of Fig.~\ref{fig:design}c we retain
designs with $\varepsilon\le1.07$ at the birth, within a few percent of the
largest anharmonicity tested directly above; these cover
$\Omega^\ast\in[0.83,1.61]$. Two designs passed directly from the linear solver
to harmonic balance produce the prescribed topology, and a third pair tests the
classifier at the type boundary itself, where $|a_\Omega|$ is $33$ and $71$
times smaller (End Matter).

The two small parameters bound different things, and only one is cut here: the
locus continues to $\Omega^\ast\simeq2.5$, where $\varepsilon$ reaches $3.7$
but $\eta_3$ is still $0.017$. They flag opposite ends of it, $\eta_3$ peaking
at $0.037$ at the low-frequency lips end that the $\varepsilon$ cut retains.
Neither is a global error bound, and the error in a fitted network parameter
involves a third factor, the local sensitivity $b_\kappa$~\cite{SM}.

Tunable coupling and experimental continuation provide practical routes for
testing the procedure~\cite{SamaniEtAl2026,SieberEtAl2008,NevilleEtAl2018}.
Cusps and hidden multistable states have both been resolved experimentally~\cite{UkaZhao2026,ZhangEtAl2026}.

The measured receptance that the procedure consumes tolerates modest sampling
noise. Relocating the birth from synthetic samples alone, at a relative
complex rms noise of $10^{-3}$, recovers $\kappa^\ast$ to
$2\times10^{-5}$--$1.2\times10^{-4}$, below the $3.6\times10^{-4}$ the
fundamental-harmonic reduction itself carries here; sweep resolution matters
much less than noise. That bounds sensitivity to noise, not to model error,
which the Supplement treats separately~\cite{SM}.

A companion paper~\cite{Sarkar2026Long} resolves the cusp contour into an
active pole and its background: a closed-form participation threshold for the
birth and its scaling with damping and mode separation, a background-sign rule
for the type and the ensemble that bounds its validity, the pole-resolved
ladder of births in multimode networks, nanomechanical scaling, and a
superconducting design estimate.

The construction relies on a specific nonlinear geometry. The nonlinear force
must act along one coordinate so that its contribution to
\eqref{eq:combbalance} is rank one; a second nonlinear coordinate or a
cross-cubic term breaks the scalar closure. The cubic coefficient is real, so
nonlinear dissipation is excluded. The result concerns periodic responses and
their period-one folds; subharmonic or quasiperiodic attractors may coexist
without moving those folds.

Coexisting amplitudes away from a cusp, and the width of the bistable window,
still require the nonlinear response. The cusp itself does not. The double-root
condition pins $|A_1|$ there, so once $\beta_1$ is known the same reduction
also fixes the cusp drive scale: at the benchmark cusp the predicted $F_c$
agrees with harmonic balance to $0.2\%$~\cite{SM}. In the examples above the comb gain
sets the scale of the residual and the first correction removes most of it,
though a nearby resonance at $3\Omega$, or proximity to the type boundary, can
make that correction large.
Once the localized cubic form and its sign are specified, the linear receptance
fixes the fundamental-harmonic location and local geometry of cusp-pair births,
so a small-signal sweep reports where the bistable band will reorganize, a gap
opening or a lens appearing, before any nonlinear continuation is run.

\begin{acknowledgments}
I thank Professor Axel Gro\ss{} (Universit\"at Ulm) and Professor Pranab
Sarkar (Visva-Bharati University) for their guidance and for many valuable
discussions. ChatGPT (free version) and Anthropic Claude were used for
language editing, and Claude additionally for drafting analysis and
figure-rendering scripts; no figure image was generated or altered by a model.
All derivations and reported numbers were regenerated and checked by the author
from the deposited scripts. I acknowledge support from the
state of Baden-W\"urttemberg through bwHPC and from the German Research
Foundation (DFG) through grant INST 40/575-1 FUGG (bwForCluster JUSTUS~2), as
well as from the Dr.~Barbara Mez-Starck Foundation.
\end{acknowledgments}

\textit{Data availability.}---The code, data, and figure scripts that support
the findings of this Letter are openly available in Zenodo at
\href{https://doi.org/10.5281/zenodo.22672005}{doi:10.5281/zenodo.22672005}. The
archive is shared with the companion paper~\cite{Sarkar2026Long}, the two
resting on one code base; its script-to-claim map (\texttt{reproduce.md})
labels each entry by the manuscript it supports.

\bibliographystyle{apsrev4-2}
\bibliography{refs}

\onecolumngrid
\vspace{10pt}
\begin{center}\rule{0.45\textwidth}{0.4pt}\end{center}
\vspace{2pt}
\begin{center}\textbf{END MATTER}\end{center}
%\twocolumngrid
\vspace{4pt}

\textit{Closure, passivity, and the Floquet comb.}---The drive may act on a
different coordinate $d$. In that case the effective forcing in
\eqref{eq:combclosure} is $FH/G$, with $H\eqdef e_1^TZ^{-1}d$ the transfer
receptance (written $H$ in the companion as well); this rescales the drive
but
leaves the cusp conditions unchanged provided $H\neq0$ at the cusp. For the
collocated receptance, passivity selects the branch in
\eqref{eq:phasecontour}. Write $Z=A+iB$ with $A=K-\Omega^2M$ and $B=\Omega C$,
and let $Zw=e_1$. Then $\operatorname{Im}\bar G=w^H B w\ge0$ for $C\succeq0$.
Hence $\operatorname{Im}G\le0$ for $\Omega>0$ and
$\arg G\in[-180^\circ,0^\circ]$, excluding the $+150^\circ$ and $+30^\circ$
levels.

The real-frequency closure requires only a linear time-invariant environment: a
more general damping law changes the dynamic stiffness sampled at each harmonic,
not the rank-one reduction. The continuation below is written for viscous
damping. The rank-one structure also reduces the Floquet problem.
Linearization about a periodic orbit $x^\ast(t)$ gives
$M\ddot{\delta x}+C\dot{\delta x}+K\delta x+p(t)\delta x_1e_1=0$ with
$p=3\beta_1x_1^{\ast2}$. With
$\delta x=e^{\lambda t}\sum_n u^{(n)}e^{in\Omega t}$, projection onto the
nonlinear coordinate gives
\begin{equation}
u_1^{(n)} + \hat G(\lambda + in\Omega)\sum_k p^{(k)} u_1^{(n-k)} = 0,
\end{equation}
where $\hat G(s)=e_1^T(Ms^2+Cs+K)^{-1}e_1$ is the viscous continuation of $G$
off the real-frequency axis. Fold conditions sample $G$ at real harmonics;
Floquet exponents sample $\hat G$ at $s=\lambda+in\Omega$. In an experiment,
this continuation would require a fitted causal model rather than the measured
real-frequency data alone.

\begin{figure}[!tbp]
  \centering
  \includegraphics[width=\columnwidth]{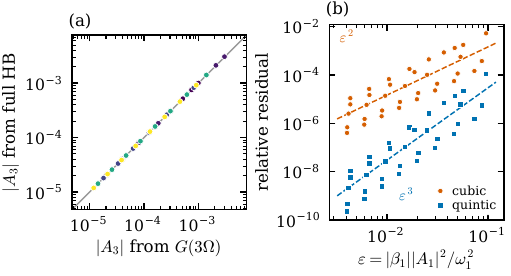}
  \caption{Numerical checks of the comb closure against converged
    multi-harmonic balance at five drive frequencies,
    $\Omega=1.05$--$1.30$. (a)~Third harmonic predicted from the linear
    receptance at $3\Omega$ and the harmonic-balance value of $A_1$, versus the
    harmonic-balance $A_3$; the line is $y=x$. (b)~Relative residual of the
    cubic and quintic reduced equations versus $\varepsilon$. The
    $\Omega$-dependent prefactor $|G(3\Omega)/G(\Omega)|$ varies tenfold, so
    exponents are fitted separately at each $\Omega$. Fits use
    $\varepsilon<0.10$; one of $36$ points lies outside this asymptotic window
    and is excluded. No quintic residual falls below the $10^{-12}$ roundoff
    cutoff.}
  \label{fig:validation}
\end{figure}

\textit{Numerical checks of the comb closure.}---As a solver check,
converged nine-harmonic balance preserves the exact cubic
rescaling to $1.4\times10^{-16}$ in the harmonic amplitudes over
$\beta_1\in[0.05,1.2]$. The next two tests add no fitted parameters
(Fig.~\ref{fig:validation}). For the first, take $A_1$ from harmonic balance and
use the linear receptance at $3\Omega$ to predict $A_3$ through
\eqref{eq:combclosure}. The predicted third harmonic agrees across $2.4$
decades with a median error of $0.2\%$. Second, adding the quintic term in
\eqref{eq:quintic} reduces the residual of the reduced equation by a median
factor of $356$. At fixed $\Omega$, the cubic and quintic residuals scale as
$\varepsilon^2$ and $\varepsilon^3$, with fitted exponents $2.00$--$2.20$ and
$3.00$--$3.20$ over the five frequencies examined.

\textit{Correction near the type boundary.}---Carrying \eqref{eq:quintic} through the
fold condition gives the corrected classifier
$\tilde a_\Omega=a_\Omega+\delta a_\Omega$~\cite{SM}. At a cusp,
$\tfrac34\beta_1|A_1|^2=-\tfrac23\rho$ for any $|\beta_1|$. Combined with the
coefficient ratio in \eqref{eq:quintic}, this makes the third-harmonic term
relative to the fundamental balance
$(\sqrt3/9)|G(3\Omega)/G(\Omega)|$. At this order the small parameter is the
linear comb ratio.

Away from the type boundary the correction is small and tracks the comb gain:
over $\omega_2 \in [0.95,\,1.25]$, $|\delta a_\Omega|/|a_\Omega|$ runs from
$0.007$ at $\omega_2 = 1.06$ to $0.028$ at $\omega_2 = 1.25$, and equals $0.45$
to $0.67$ times $|G(3\Omega)/G(\Omega)|$, which itself spans
$1.5$ to $5.4\times10^{-2}$. Within this corrected model the ratio is exactly
$\beta_1$-independent, as \eqref{eq:quintic} requires: at $\omega_2 = 1.25$,
$\delta a_\Omega = -36.51$ against $a_\Omega = +1315$ for every $\beta_1$ in
$[0.05,\,1.2]$.

The type boundary is a fold of the birth locus itself. Just below
$\omega_2\simeq1.049$, the two-mode family has one birth near
$\kappa\approx0.18$. As $\omega_2$ increases, two additional births appear near
$\kappa=0.033$, one beaks and one lips, with $a_\Omega$ changing sign between
them. This fold of the birth locus produces the overlap of the two design types
in Fig.~\ref{fig:design}c.

Just above this threshold the two new births are close. At
$\omega_2=1.0494$ they are $6\times10^{-5}$ apart in $\kappa$, with
$a_\Omega=+4.89$ and $-4.30$; at $\omega_2=1.0502$ the separation is
$3.9\times10^{-4}$, with $a_\Omega=+9.55$ and $-7.51$. The corrected values are
$+5.17/-4.51$ and $+9.67/-7.58$, respectively, so
$|\delta a_\Omega|\approx0.07$--$0.28$. Using
$|\partial_\kappa a_\Omega|\sim4\times10^4$--$1.5\times10^5$, the condition
$|\delta a_\Omega|\ge|a_\Omega|$ is confined to about
$2\times10^{-6}$ in $\kappa$, six parts in $10^5$ of the local coupling. This
width is a second-order estimate for this family; a resonance near $3\Omega$
would increase the correction and broaden this near-boundary region.

The classification itself was tested there. At $\omega_2=1.06$ the family
carries two births of opposite type---a beaks at
$(\Omega^\ast,\kappa^\ast)=(1.0785,0.04100)$ with $a_\Omega=+40.0$, and a
lips at $(1.0611,0.05922)$ with $a_\Omega=-18.6$---both at
$\varepsilon\simeq0.1$ and $\eta_3\simeq3\times10^{-3}$, so a failure here
would be the classifier's and not the reduction's. Multi-harmonic balance
reproduces both. Across the beaks point the connected band $[1.046,1.110]$ at
$\kappa=0.0360$ splits into $[1.050,1.070]$ and $[1.092,1.110]$ at
$\kappa=0.0460$, a gap straddling $\Omega^\ast$ whose edges match the
predicted cusps $1.0703$ and $1.0909$ to the $2\times10^{-3}$ scan step.
At the lips point $b_\kappa<0$, the opposite sign to the
$\omega_2=0.95$ case of Fig.~\ref{fig:design}b, so the oval lies below
$\kappa^\ast$: at $\kappa=0.0542$ an isolated band $[1.0568,1.0647]$ sits
clear of a second band beginning at $1.1023$, against predicted cusps
$1.0551$, $1.0658$ and $1.1017$; the scan window ends at $1.130$, so the upper
band's far edge is not resolved here. The oval has gone by $\kappa=0.0642$. It
is the sub-wedge left by the beaks birth, contracting and annihilating, not a
lens appearing from an empty neighborhood. Its width follows the predicted
$c_\ast\sqrt\mu$ law, with the threshold displaced about $0.7\%$ upward in
$\kappa$~\cite{SM}. The classifier
therefore assigns the correct topology on both sides of a boundary it places
inside a window narrower than the resolution of the test. Whether the third-harmonic correction displaces that boundary is not
resolved here, and is not claimed.

\textit{Locating the birth.}---The codimension-two birth can be located
without the $3/2$ exponent at all. We
bisect directly on the coupling at which a gap first appears in the bistable
band of the harmonic-balance solution. Two runs of this test were configured
independently: a single-setting run, and one repeated over three frequency
samplings, three scan windows, two arclength steps and $N_{\rm HARM}=3$ and
$4$. The test is one-sided---a detected gap places $\kappa^\ast$ firmly below
the tested coupling, whereas an undetected gap places it above only to the
frequency-scan detection floor, $\lesssim10^{-6}$ in $\kappa$---so the two are
combined by intersection, giving
$\kappa^\ast\in[0.1193188,0.1193203]$~\cite{SM}. At fixed coupling the
verdict is truncation independent, returning the same answer at
$N_{\rm HARM}=3$, $5$, $7$ and $9$. The independent
$(\Delta F)^{2/3}$ extrapolation gives $\kappa^\ast=0.11907$, and the first
third-harmonic corrected cusp condition gives $\kappa^\ast=0.1192884$
(Supplement), reducing the discrepancy from the fundamental linear prediction
by a factor of $11.6$, between $11.3$ and $12.3$ across the bracket and its
soft edge. The
$(\Delta F)^{2/3}$ value sits $2.5\times10^{-4}$ out because it inherits the fit-window systematic of the Supplement, not a
truncation error. Both are consistency checks on an event the bisection located
without either estimator.

\textit{Topology and stability.}---For the hub of Fig.~\ref{fig:design}a, the
bistable band below the birth is connected throughout the scanned range,
$\Omega\in[1.11,\,1.28]$. Above the birth, harmonic balance finds a gap from
$1.175$ to $1.223$, bracketing the predicted cusp pair at $1.179$ and $1.218$
within one scan step of $8\times10^{-3}$. The gap widens with coupling: at
$\kappa=0.14$ it runs from $1.183$ to $1.256$, against a predicted pair at
$1.187$ and $1.253$.

For the absorber of Fig.~\ref{fig:design}b, no bistability is found over
$\Omega\in[1.001,\,1.069]$ below the birth. Above it an isolated lens appears;
at $\kappa=0.333$ harmonic balance resolves the lens over
$\Omega\in[1.01982,\,1.06454]$, compared with the predicted interval
$[1.01941,\,1.06495]$. The frequency step is $1.9\times10^{-3}$, so the
comparison is already at the scan resolution: each edge differs by $0.22$ of a
step and the width by $0.44$ of a step. The resolved edges have
$\arg G=-150.12^\circ$ and $-150.13^\circ$. We use the resolved band rather than an extrapolated tip
because the $(\Delta F)^{2/3}$ estimator fails for this family~\cite{SM}.

Reconstructing the coexisting orbits from their harmonic-balance coefficients
and integrating the monodromy over one drive period gives Floquet spectral
radii of $0.91$--$0.92$ on each hub sub-wedge. For the lens the outer branches
have radii $0.963$ and $0.982$, the intervening one $1.011$.

\end{document}